\documentclass[iop,apj]{emulateapj}

\usepackage{natbib}
\usepackage{amsmath}
\usepackage{multirow}
\usepackage{float}
\usepackage{color}
\usepackage{afterpage}
\usepackage[abs]{overpic}
\usepackage{CJK}
\usepackage{hyperref}
\usepackage{rotating}

\definecolor{Red}{rgb}{1,0,0}
\definecolor{Blue}{rgb}{0,0,1}
\definecolor{Green}{rgb}{0,1,0}
\definecolor{magenta}{rgb}{1,0,.6}
\definecolor{lightblue}{rgb}{0,.5,1}
\definecolor{lightpurple}{rgb}{.6,.4,1}
\definecolor{gold}{rgb}{.6,.5,0}
\definecolor{orange}{rgb}{1,0.4,0}
\definecolor{hotpink}{rgb}{1,0,0.5}
\definecolor{newcolor2}{rgb}{.5,.3,.5}
\definecolor{newcolor}{rgb}{0,.3,1}
\definecolor{newcolor3}{rgb}{1,0,.35}
\definecolor{darkgreen1}{rgb}{0, .35, 0}
\definecolor{darkgreen}{rgb}{0, .6, 0}
\definecolor{darkred}{rgb}{.75,0,0}

\def \hmsol     {h^{-1}{\rm\ M}_\odot}
\def \hMpc      {h^{-1}{\rm\ Mpc}}

\newcommand{\mpccube}{\ensuremath{\, h^{-3}\,\mathrm{Mpc}^3 }}

\newcommand{\lya}{Ly$\alpha$}

\newcommand{\lyaf}{Ly$\alpha$ forest}

\newcommand{\beq}{\begin{equation}}
\newcommand{\eeq}{\end{equation}}
\newcommand{\bc}{\begin{center}}
\newcommand{\ec}{\end{center}}
\newcommand{\bfig}{\begin{figure}}
\newcommand{\efig}{\end{figure}}

\newcommand{\sqdeg}{\mathrm{deg}^{2}}

\newcommand{\dperp}{\ensuremath{d_\perp}}
\newcommand{\dsmall}{\textit{d25}\ }
\newcommand{\dmid}{\textit{d40}\ }
\newcommand{\Ddm}{\ensuremath{\Delta_\mathrm{dm}}}

\slugcomment{Accepted by ApJ}
\shorttitle{Characterizing $z \sim 2.5$ Cosmic Web With \lya\ Forest Tomography}
\shortauthors{Lee \& White}

\begin{document}

\title{ 
Revealing the $\lowercase{z} \sim 2.5$ Cosmic Web With 3D \lya\ Forest Tomography: \\
A Deformation Tensor Approach}
\author{Khee-Gan Lee\altaffilmark{1,3} \& 
Martin White\altaffilmark{1,2}
}
\altaffiltext{1}{Lawrence Berkeley National Laboratory, 1 Cyclotron Rd., Berkeley, CA 94720, USA}
\altaffiltext{2}{Departments of Physics and Astronomy, University of California at Berkeley, New Campbell Hall,
Berkeley, CA 94720, USA}
\altaffiltext{3}{Hubble Fellow}
\email{kglee@lbl.gov}

\begin{abstract}
Studies of cosmological objects should take into account their positions within the cosmic web of large-scale structure. Unfortunately, the cosmic web has only been extensively mapped at low-redshifts ($z<1$), using galaxy redshifts as tracers of the underlying density field.  At $z>1$, the required galaxy densities are inaccessible for the foreseeable future, but 3D reconstructions of \lyaf\ absorption in closely-separated background QSOs and star-forming galaxies already offer a detailed window into $z\sim2-3$ large-scale structure.
We quantify the utility of such maps for studying the cosmic web by using realistic $z=2.5$ \lyaf\ simulations 
matched to observational properties of upcoming surveys.
A deformation tensor-based analysis is used to classify voids, sheets, filaments and nodes in the flux, which is
compared to those determined from the underlying dark matter field.  We find an extremely good correspondence, with $70\%$ of the volume in the flux maps correctly classified relative to the dark matter web, and 99\% classified to within 1 eigenvalue.  This compares favorably to the performance of galaxy-based classifiers with even the highest galaxy densities from low-redshift surveys.  We find that narrow survey geometries can degrade the cosmic web recovery unless the survey is $\gtrsim60\,\hMpc$ or $\gtrsim1\,\mathrm{deg}$ on the sky.  
We also examine halo abundances as a function of the cosmic web, and find a clear dependence as a function of flux overdensity, but little explicit dependence on the cosmic web.
These methods will provide a new window on cosmological environments of galaxies at this very special time in galaxy formation, ``high noon'', and on overall properties of cosmological structures at this epoch.

\end{abstract}


\section{Introduction}

The process of gravitational instability from Gaussian initial conditions generically produces a beaded, filamentary network within which dark matter halos and associated cosmological objects form.  As the Universe evolves, matter flows out of voids and through sheets before accreting along filaments into nodes or halos.    This distinctive pattern of voids, sheets, filaments and nodes is known as the ``cosmic web'' \citep{zeldovich:1982,
shandarin:1983,einasto:1984,bond:1996,colberg:2007} and forms the context within which structure formation takes place, and populations of cosmological objects form and evolve.
The cosmic web has been seen in the distribution of galaxies in the local Universe, and it is a robust prediction of cosmological simulations dating back many years.
However, an understanding of the evolution of objects within the context of this cosmic web remains a major, unsolved problem in observational cosmology.

It has become increasingly apparent that the ``environment'' of galaxies plays an important role in their morphology, evolution, the evolution of their stellar populations and their level of nuclear activity 
\citep[e.g.,][]{dressler:1980,kauffmann:2004,blanton:2005,blanton:2007}.
It is still an open question, especially at high redshift, to what degree these effects depend on the immediate neighbors of a galaxy or extend to large-scale tidal field and the rich structures in which these objects are embedded.
At low redshift, galaxy color and morphology depend primarily on local density \citep{yan:2013} as does the galaxy luminosity function \citep[e.g.~most recently:][]{eardley:2015} although filamentarity has a direct effect on galaxy formation \citep{guo:2014} and possibly star-formation \citep{snedden:2016}.  There is evidence that dwarf galaxy evolution may be affected by the cosmic web \citep{benitez-llambay:2013}, as may star-formation in galaxies falling into clusters along filaments \citep{porter:2008}.  The properties of dark matter halos are seen to depend on their locations within the web \citep{hahn:2007,ludlow:2011}; with low mass void galaxies being younger and halo spins\footnote{To the extent that this affects the intrinsic shapes of galaxies, this could be a source of systematic error for weak-lensing 
cosmic shear analyses.} of sheet galaxies lying preferentially within the plane of symmetry of the mass distribution. \citet{dubois:2014}, using hydrodynamic simulations, predict that spins of red and blue galaxies 
should align perpendicularly and parallel, respectively, with filaments at high redshift ($z\gtrsim 1$).
There is currently very little observational information about how galaxy properties depend upon the web at redshifts near the peak of cosmic star-formation ($z\sim 1.5-3$) or beyond, although hydrodynamical simulations are beginning to deliver predictions at this
epoch, e.g. \citet{feldmann:2016}.

The cosmic web is most frequently studied in the distribution of optical galaxies from 
spectroscopic redshift surveys.
In the past decade, studies of the nearby universe ($z < 0.2$)
\citep{bond:2010,alpaslan:2015,leclercq:2015}
have reached a high degree of maturity, characterizing the web and its denizens in some detail 
(see \citealt{alpaslan:2015} for a recent, low-$z$, example with references to the earlier literature).
For this kind of analysis a high number density of galaxies with spectroscopic redshifts is required, 
which becomes increasingly expensive at higher-redshift.
This makes it difficult to accumulate a sample of sub-$L_\star$ galaxies (with $n^{-1/3} \sim\,\mathrm{Mpc}$) over cosmologically representative volumes that can simultaneously resolve cosmic web components, which characteristically span of tens of Mpc \citep[e.g.][]{colberg:2007,stark:2015,stark:2015a}
As an example, the ongoing VIPERS survey \citep{guzzo:2014} on the $8.2\,$m VLT represents the current state-of-the-art in terms of galaxy redshift surveys that could map the cosmic web at higher redshifts.  With a mean redshift of $\langle z \rangle \sim 0.7$, VIPERS required $>400\,$hrs of observing time to achieve a number density of $\bar{n}\simeq 2\times10^{-3}\,h^{3}\mathrm{Mpc}^{-3}$, or a typical spatial resolution of $\bar{n}^{-1/3}\simeq 8\,h^{-1}$Mpc, which is marginal for studying the cosmic web.
This route to mapping the cosmic web at $z>1$, over cosmological volumes, would thus appear to be a task only feasible with future $30\,$m-class telescopes, and even then requiring massive investments of telescope time.

However, another tracer of the dark matter is the intergalactic medium (IGM).
We have recently shown \citep{lee:2014a, lee:2016} that observations of the IGM \lyaf\ absorption in closely-separated background QSOs and star-forming galaxies (LBGs) at $z\sim 2-3$ can allow 3D reconstructions of the IGM absorption that resolve scales of several Mpc. \citet{lee:2014} showed that applying Wiener-filter reconstruction methods to simulated sightlines, matched to upcoming data in terms of transverse separation, resolution, and signal-to-noise, yield maps that agree visually with the underlying matter density field smoothed on similar scales.  \citet{stark:2015} and \citet{stark:2015a} (hereafter S15a and S15b, respectively) showed that such maps can be used to detect large samples of high-redshift galaxy protoclusters and voids, respectively. In both cases, simulated IGM tomographic maps were found to recover these extended structures with good purity and completeness.

This IGM reconstruction is being pioneered by the CLAMATO 
survey\footnote{\url{http://clamato.lbl.gov}}, conducted on the LRIS spectrograph \citep{oke:1995} on the $10.3\,$m diameter Keck-I telescope.
CLAMATO, which is completing its pilot phase at the time of writing, aims to map the IGM with an effective spatial resolution of $\sim 3\,\hMpc$ over a $\sim1\,\sqdeg$ area at $2.2<z<2.5$, yielding a total, comoving, survey volume of $\sim10^6\,\mpccube$.
In addition, in the near future, the Subaru Prime-Focus Spectrograph (PFS) -- a wide-field and massively-multiplexed spectrograph capable of targeting over 2000 objects within a $\sim 1\sqdeg$ field-of-view -- will commence operation \citep{takada:2014}. Since PFS samples a smaller telescope aperture ($8.2\,$m diameter) and is less blue-sensitive than Keck-LRIS, it cannot reach similar depths and sightline densities as CLAMATO but will be especially well suited to tomographic mapping over much larger areas ($\sim 20\,\sqdeg$) and volumes ($\sim 10^8\,\hMpc$), albeit with a coarser effective spatial resolution ($\sim 5\,\hMpc$).
 
Voids and protoclusters roughly correspond to the two limiting types of structures in the cosmic web, as well as being the largest spatially coherent structures.  Their detection and characterization thus form a natural starting point for surveys such as CLAMATO, which explains the focus of S15a and S15b.  Here we analyze the same $z=2.5$ simulated IGM tomographic maps as S15a, S15b to see whether maps such as those returned by CLAMATO, or a future PFS survey, could further classify the cosmic web, i.e.\ into voids, sheets, filaments and nodes. As we shall see, there is a remarkable agreement between the structures recovered from realistic IGM maps, and those in the underlying dark matter (DM) field.  We will also conduct a preliminary investigation into the variation of halo abundances in different overdensity and cosmic environments of the high-redshift cosmic web, which is intended to motivate
more detailed in this direction using either hydrodynamical simulations or semi-analytic galaxy formation
models.

This paper is not the first attempt at analyzing the 3D cosmic web in IGM simulations: \citet{caucci:2008} had previously performed an analysis of cosmic web filaments in simulated IGM tomographic maps using the Skeleton algorithm \citep{sousbie:2009}.
They considered noiseless mock spectra with no observational effects taken into account except random sightline sampling.  We take the next steps by including further observational effects tailored to existing and near-future surveys, with an eye to the recovery of the $z\sim 2.5$ cosmic web in the observable IGM (and its relation to the web in dark matter).

\section{Simulations}
\label{sec:sims}

In this paper, we use the same, numerical, $N$-body simulations described in S15a and S15b. These are dark matter-only simulations with $2560^3$ equal-mass ($8.6\times 10^7\,\hmsol$) particles in a periodic box with $L=256\,\hMpc$ on a side. The comoving volume is large enough to encompass the largest cosmic structures (see e.g.~S15b), while the mass resolution is sufficient to model the \lya\ forest absorption at $z\sim 2-3$ using the ``fluctuating Gunn-Peterson approximation'' \citep{croft:1998, meiksin:2001, white:2010, rorai:2013,mcquinn:2015}. 

The simulation adopted a flat $\Lambda$CDM cosmology with $\Omega_{\mathrm{m}}\approx 0.31$, 
$\Omega_{\rm b} h^2 \approx 0.022$, $h = 0.6777$, $n_s = 0.9611$ and $\sigma_8 = 0.83$, consistent with \citet{planck-collaboration:2014}. The initial conditions were generated using second-order Lagrangian perturbation theory at $z=150$, with subsequent evolution of the particle positions and velocities solved using the TreePM code \citep{white:2002}. The $z=2.5$ particle positions and velocities were used to generate \lya\ forest absorption skewers using the fluctuating Gunn-Peterson approximation, assuming a power-law IGM temperature-density relationship, $T = T_0 (\rho/\bar{\rho})^{\gamma-1}$ with slope $\gamma=1.6$ 
\citep[consistent with observational constraints, e.g.,][]{rudie:2012, lee:2015}. 

The overall size of the simulation box ($L=256\,\hMpc$) is slightly small for a dark matter-only
simulation, but we are constrained by the need to simultaneously resolve the 
$\sim 100\,\mathrm{kpc}$ Jeans' scale of the IGM \citep[see, e.g.,][]{rorai:2013,sorini:2016}. However, the characteristic scale of cosmic web structures
typically span tens of Mpc at $z\sim 0$ \citep[e.g.,][]{colberg:2007} and slightly smaller
at high-redshift \citep{stark:2015,stark:2015a}, therefore we believe our simulation
to be sufficiently large for our current purpose. 
Moreover, we will be using Fourier transforms to perform our cosmic web analysis; this
would not introduce errors on our fully periodic box, although as we shall see 
non-periodic sub-volumes do indeed lead to errors.

To construct mock observations, we generated 409600 absorption skewers with a transverse
spacing of $0.4\,\hMpc$ in the $xy$ plane. We work throughout in terms of the \lyaf\ flux fluctuation 
\beq \delta_F=F/\langle F \rangle -1, \eeq
where $F=\exp(-\tau)$ is the transmitted flux fraction and $\tau$ is the \lya\ optical depth.
For brevity, we will use `flux' to refer to $\delta_F$.

For this analysis, we use the publicly-available simulation products\footnote{\url{http://tinyurl.com/lya-tomography-sim-data}} described in S15a. 
Since we aim to compare the potentially observable IGM field with the underlying cosmic web in the large-scale structure, we will use both the mock IGM tomographic maps and the gridded, DM density field from these simulations. For the former, we use primarily the {\it d25} flux maps, which are Wiener-filtered reconstructions of randomly-selected simulation skewers using the method described in S15a, specifically designed to match the properties of the CLAMATO pilot data of \citet{lee:2014a}. 
These maps assumed an average sightline separation of $\langle \dperp \rangle = 2.5\,\hMpc$,
with line-of-sight Gaussian smoothing equivalent to a spectral resolution of 
$R\equiv \lambda / \Delta \lambda = 1100$ with pixel noise added as described in S15a.
The resulting tomographic maps, binned on a $1\,\hMpc$ grid, should accurately reflect the 
main properties of CLAMATO observations.

To compare with a hypothetical survey that might be carried out with the upcoming Subaru PFS instrument, 
we will also use the {\it d40} map created with $\dperp=4\,\hMpc$ sightline spacing. This decreased sightline spacing reflects a shallower ($g \sim 24$ rather than $g \sim 24.5$ in CLAMATO) but wider survey ($\sim 20\,\sqdeg$ rather than $\sim 1\,\sqdeg$) that might be better suited to the lower throughput but much 
wider field-of-view of the PFS.

To validate the cosmic web recovery from the tomographic \lya\ flux maps, we will adopt as underlying `truth' the $z=2.5$ redshift-space dark matter overdensity field $\Ddm = 1+\delta_{\rm dm} = \rho_m/\langle \rho_m \rangle$ binned into a regular grid with $1\,\hMpc$ cell size
and then Gaussian-smoothed to match each tomographic map.

\section{Cosmic Web Classification Scheme}

With the rich and complex of the cosmic web, there are numerous methods by which one can classify the elements of the cosmic web. These range from techniques utilizing minimal-spanning trees
\citep{barrow:1985,alpaslan:2014}, the Candy model \citep{stoica:2005, stoica:2010}, Morse theory-based
`skeleton' methods \citep{colombi:2000,novikov:2006,sousbie:2009,sousbie:2011}, tessellation methods
\citep{gonzalez:2010}, multi-scale morphological methods \citep{aragon-calvo:2007,aragon-calvo:2010,
cautun:2013}, ridge-finding \citep{chen:2015}, and others;
\citep[see ][for a comprehensive summary]{cautun:2014}.

In this paper, it will prove convenient to use a``deformation tensor'' approach based 
on the Hessian of the gravitational potential, which is inspired by \citet{zeldovich:1970}.  
This method can be motivated by a linearization of the equations of motion around extrema \citep{hahn:2007}.  It has been applied by \citet{hahn:2007} and \citet{forero-romero:2009} to analyze the cosmic web in modern cosmological simulations, as well as to galaxy redshift survey data by \citet{eardley:2015} and we refer the reader to those papers for further discussion.

The Hessian matrix needs to be computed on a smoothed density or flux field.  
We therefore smooth our gridded fields with a Gaussian kernel whose width is chosen to be approximately $1.5\times$ the average sightline separation
of the mock data from which the tomographic flux maps are reconstructed.  
This reduces reconstruction noise from finite sightline sampling and pixel noise \citep{caucci:2008}.
Specifically we smooth by $R_G=4\,\hMpc$ for the \dsmall reconstructions ($\dperp=2.5\,\hMpc$) and $R_G=6\,\hMpc$ for the \dmid maps ($\dperp=4\,\hMpc$).
We also smooth the DM field by the same $R_G$ in order to directly compare with the smoothed tomographic map.
The $R_G=4\,\hMpc$ scale is, conveniently, the smoothing used in several recent analyses of low-redshift galaxy surveys and simulations \citep[e.g.,][]{eardley:2015}. This could, in principle, facilitate a comparison between the observed cosmic web at $z\sim 2.5$ and $z\sim 0.2$ from CLAMATO and GAMA once both data sets are available.

Note that technically these two smoothed fields should not be directly
compared with each other since the transformation from DM to the \lya\ forest does not
commute with the Gaussian smoothing, but visually the cosmic web features in the two 
smoothed fields appear to agree well. We therefore we proceed with a direct
comparison of the smoothed DM and IGM fields, and leave a more rigorous analysis
to future work.

We would like to classify the cosmic web components in both the dark matter and the IGM.
For the dark matter, the method largely follows that from \citep{forero-romero:2009}:
 after smoothing, we compute the deformation tensor of the density field, which is the Hessian of the (pseudo-)gravitational potential, $\Phi$:
\beq
T_{ij} = \frac{\partial^2 \Phi}{\partial x_i \partial x_j }. 
\eeq
This is most conveniently calculated in Fourier space, where
$\nabla^2 \tilde{\Phi} = k^2 \tilde{\Phi} =\delta_k$ (in units with $4\pi G = 1$) and therefore
\beq \label{eq:deform}
\widetilde{T}_{ij}(\vec{k}) = \frac{k_i k_j}{k^2}\ \delta_k,
\eeq
where $\delta_k$ is the Fourier-space overdensity.
This is then inverse Fourier-transformed into configuration space to obtain $T_{ij}$.

We solve for the three eigenvalues of $T_{ij}$, $\lambda_k$, at each grid point in the map. The geometric cosmic web classification is determined by the number of eigenvalues that are greater than a given threshold\footnote{By contrast the number of positive eigenvalues is the dimension of the stable manifold of the dynamical system at the fixed points.}, $\lambda_k \geq \lambda_{\rm th}$: voids, sheets, filaments and nodes, respectively, have zero, one, two, and three eigenvalues greater than the threshold.
The choice of $\lambda_{\rm th}$ is a free parameter, related to a characteristic
collapse time-scale \citep{forero-romero:2009}, that in practice primarily governs the relative 
volume-filling factors of the different components.  Many different criteria have been
considered in the literature for determining $\lambda_{\rm th}$, e.g.\ percolation
of voids and filaments \citep{forero-romero:2009} or equipartition of the
 cosmic web components \citep{eardley:2015}.
In this work, we set $\lambda_{\rm th}$ to yield a void fraction roughly similar to that from S15b, which was approximately $19\%$ and $17\%$ for the $R_G=[4, 6]\,\hMpc$ maps, respectively.

To classify positions in the IGM in terms of the cosmic web 
one should, in principle, first carry out an inversion of the flux field to estimate the
underlying DM distribution \citep[e.g.,][]{pichon:2001,caucci:2008}. 
However, 
in this paper we will instead begin with a simple \textit{ansatz}:
we directly substitute the flux Fourier modes from the mock IGM tomographic maps ($\delta_{F,k}$) for $\delta_k$ in Eq.~(\ref{eq:deform}) to estimate the IGM pseudo-deformation tensor. As high flux corresponds to low density, we reverse the sign of the eigenvalue threshold criterion for classifying voids, sheets, filaments and nodes.  Thus we count eigenvalues for which $\lambda_k \leq \lambda_{{\rm th},F}$.  Again, $\lambda_{{\rm th},F}$ is chosen to give the void fraction found by S15b.  In practice, $\lambda_{{\rm th},F} \ne \lambda_{\rm th}$.
As we shall see, even this simplistic method results in a good match
between the classification of the cosmic web elements in the dark matter and 
reconstructed flux, which can be regarded as a conservative lower limit in the
cosmic web recovery from upcoming surveys. More sophisticated methods for web recovery
should improve on our results, but we defer that to future work.

\section{Cosmic Web Recovery}

\begin{deluxetable*}{l c  c  c  c c c }
\tablecolumns{7}
\tablecaption{\label{tab:basicfrac} Volume-Filling Fractions of Cosmic Web Components}
\tablehead{
\multirow{2}{*}{Map} & Smoothing & Eigenvalue & \multicolumn{4}{c}{Volume-Filling Fraction (\%)} \\
\noalign{\vskip 0.3em}
\cline{4-7} \noalign{\vskip 0.3em} & $(\hMpc)$ & Threshold & \colhead{Voids} & \colhead{Sheets} & \colhead{Filaments} & \colhead{Nodes}
}
\startdata
Dark Matter & 4 & $\lambda_{th} \geq 0.070$ & 19.3 & 48.4 & 28.1 & 4.2 \\
\textit{d25} Flux Map & 4  & $\lambda_{th,F} \leq -0.0050$ & 19.3 & 48.7 & 28.2 & 3.8 \\[0.5em]
Dark Matter & 6 & $\lambda_{th} \geq 0.082$ & 17.5 &  45.8 &  31.5 &   5.2 \\
\textit{d40} Flux Map & 6 & $\lambda_{th,F} \leq -0.0031$ & 17.5 &  46.8  &  31.0 & 4.7
\enddata
\end{deluxetable*}

\begin{figure*}
\begin{center}
\fbox{\includegraphics[width=0.85\textwidth]{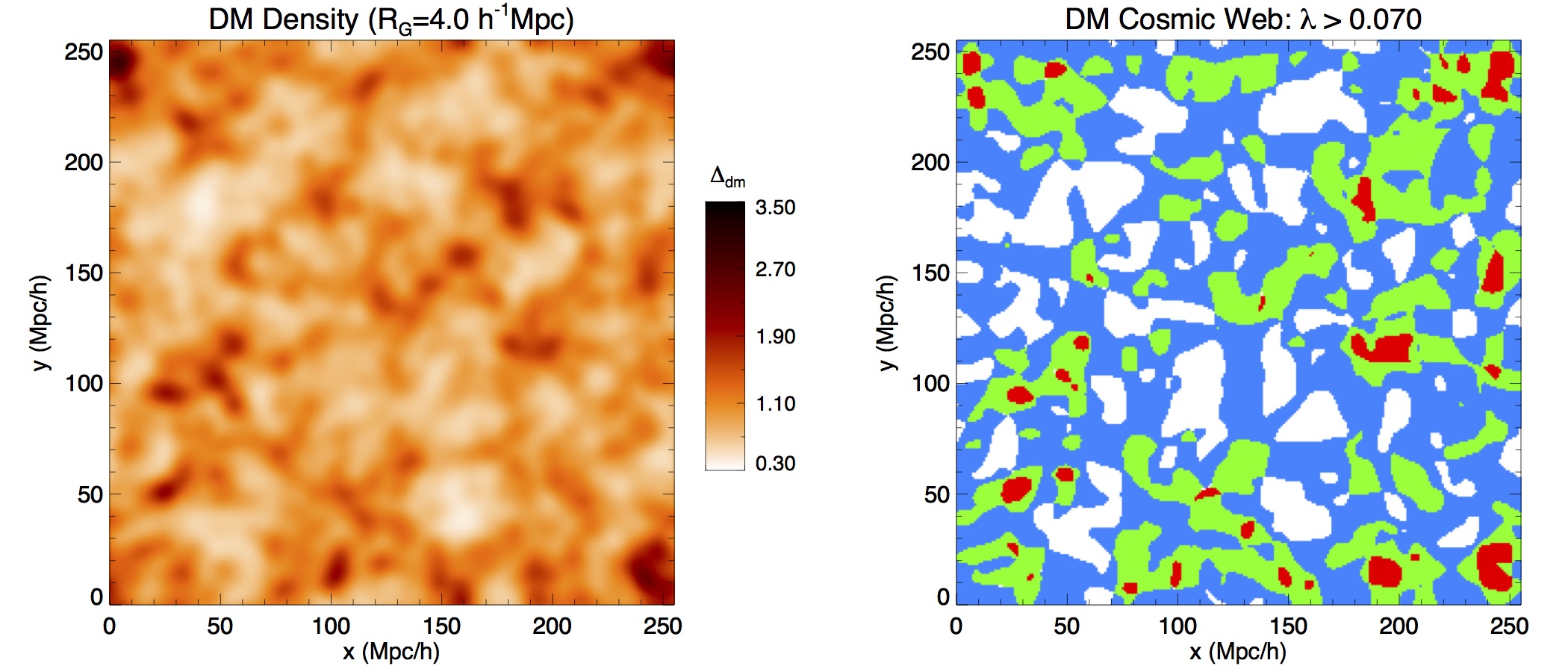}}\\[2em]
\fbox{\includegraphics[width=0.85\textwidth]{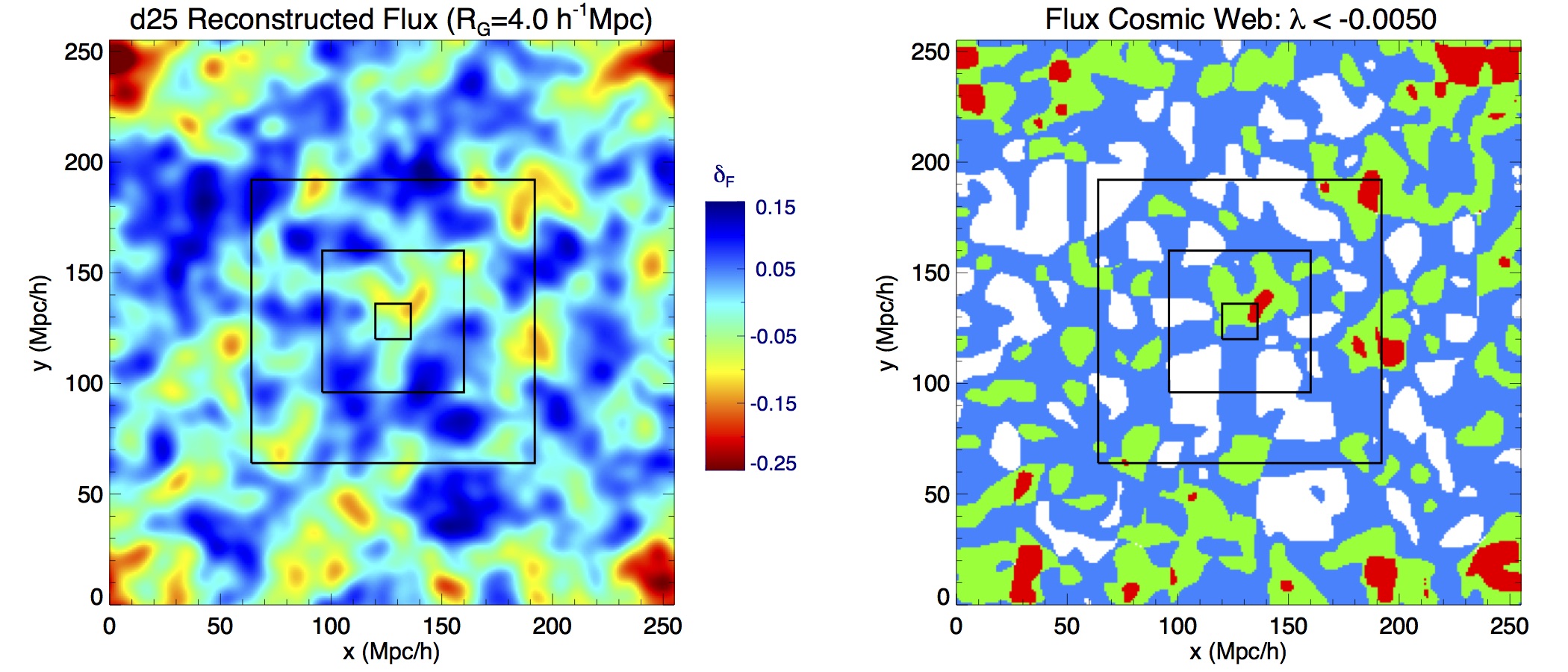}} 
\end{center}
\caption{\label{fig:cosmicweb1}
Map slices showing the simulated dark matter field (top left) and \dsmall 
mock \lya\ forest tomographic reconstruction in the same volume
(bottom left), both Gaussian-smoothed to $R_G=4\,\hMpc$. Panels at right show the corresponding cosmic
web classifications recovered by applying the deformation tensor technique to the corresponding map to the left: 
voids are in white, sheets in light-blue, filaments in green, and nodes in red.  Each slice has 
a thickness of $2\,\hMpc$ along the line-of-sight $z$-direction. The black squares overplotted on the 
flux maps (bottom row) indicate the transverse areas corresponding to $(16\,\hMpc)^2$, $(64\,\hMpc)^2$, 
and $(128\,\hMpc)^2$. The cosmic web recovered from
the tomographic flux, which includes realistic observational effects, agrees remarkably well with that from the
underlying DM field: $[15\%,69\%,15\%]$ of the volume are classified to within $[-1,0,+1]$ eigenvalues.
}
\end{figure*}

\begin{figure*}
\begin{center}
\fbox{\includegraphics[width=0.85\textwidth]{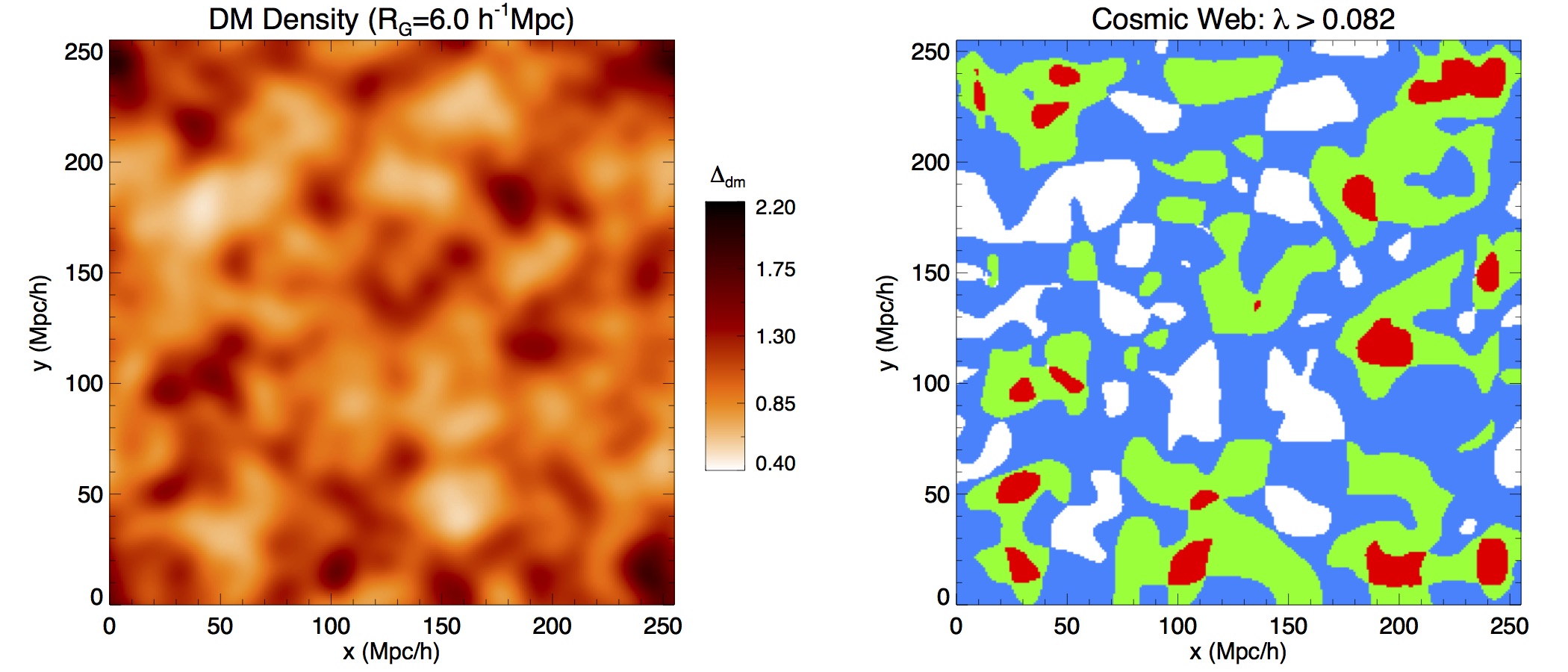}}\\[2em]
\fbox{\includegraphics[width=0.85\textwidth]{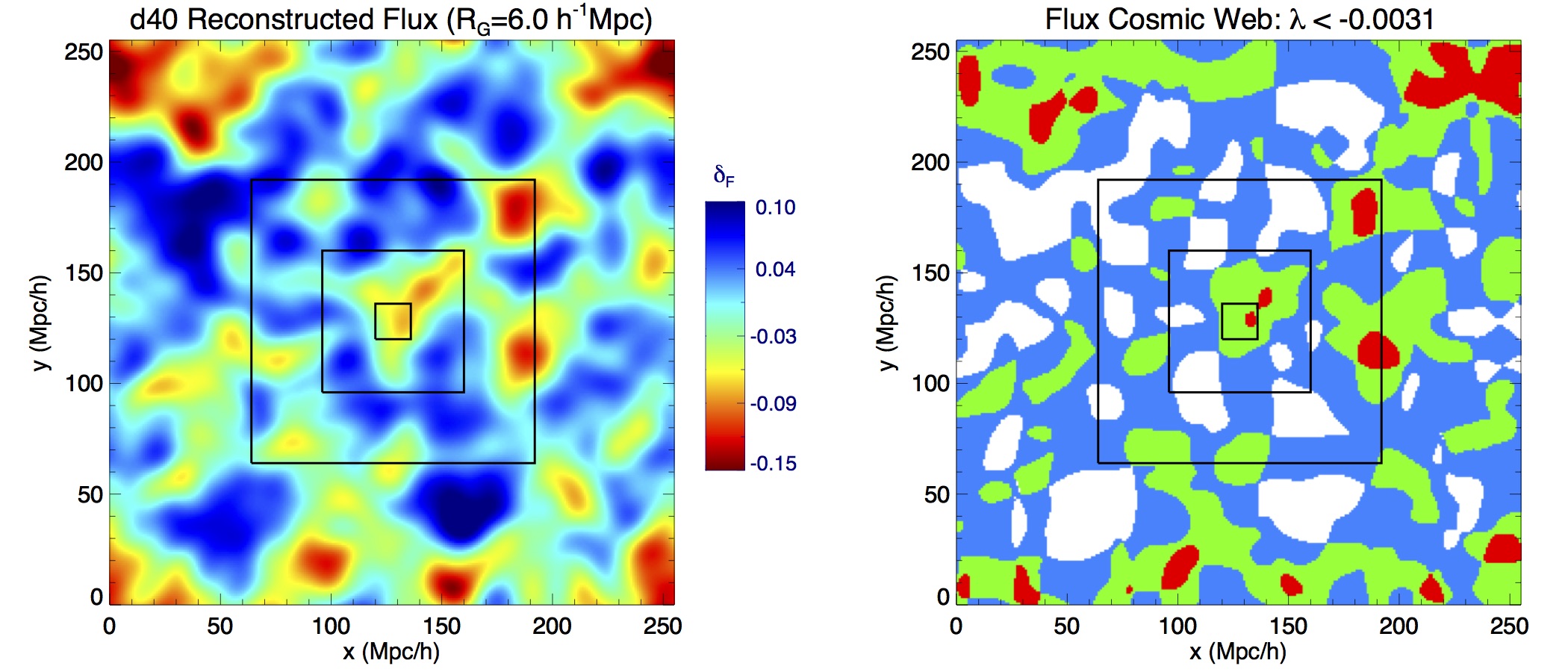}} 
\end{center}
\caption{\label{fig:cosmicweb2}
Same as Fig.~\ref{fig:cosmicweb1}, but shown for the DM field smoothed to $R_G=6\,\hMpc$ and
\dmid tomographic reconstruction.
}
\end{figure*}

The cosmic web recovered from the $R_G = 4\,\hMpc$ DM field and \dsmall
tomographic flux maps are illustrated in Fig.~\ref{fig:cosmicweb1}, with the corresponding
volume-filling fractions of voids, sheets, filaments, and nodes tabulated in Table~\ref{tab:basicfrac}.
Fig.~\ref{fig:cosmicweb2} shows the same for the $R_G=6\,\hMpc$ DM maps with the \dmid 
reconstructions.
Visually, the voids in the maps appear to trace the lowest-density regions 
while nodes trace the highest-density regions, as expected. 
Our choice of threshold also seems, by eye, to yield sheets and filaments that are
appropriately pancake- and tendril-like, respectively. 

\begin{figure*}
\includegraphics[width=0.49\textwidth]{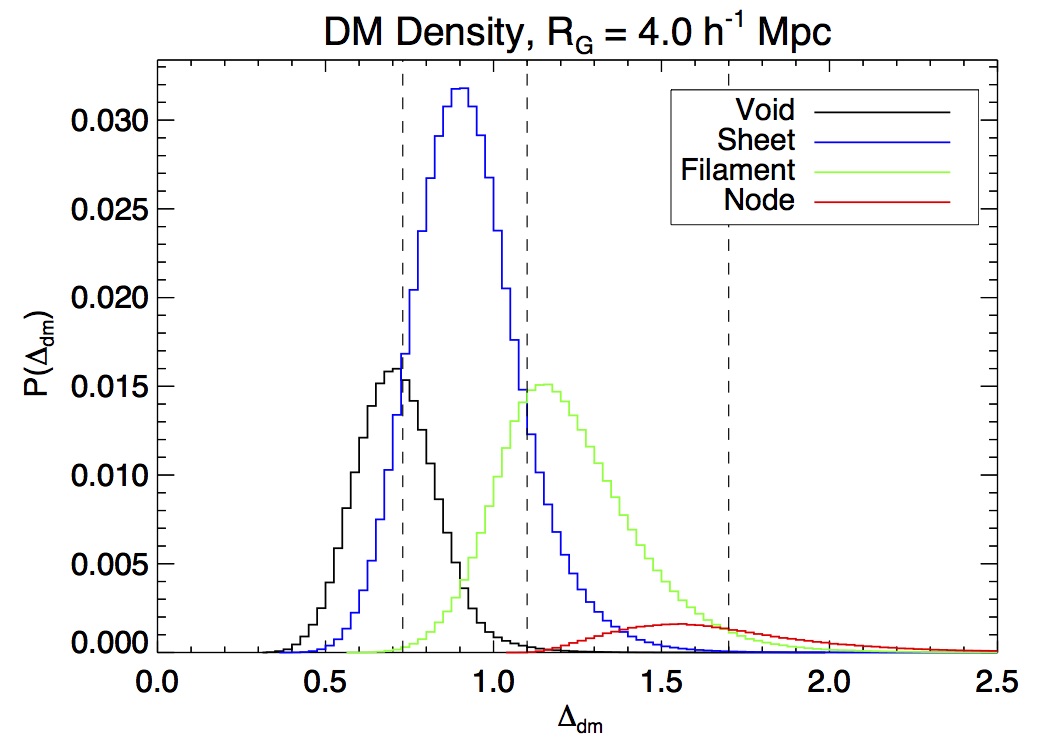}
\includegraphics[width=0.49\textwidth]{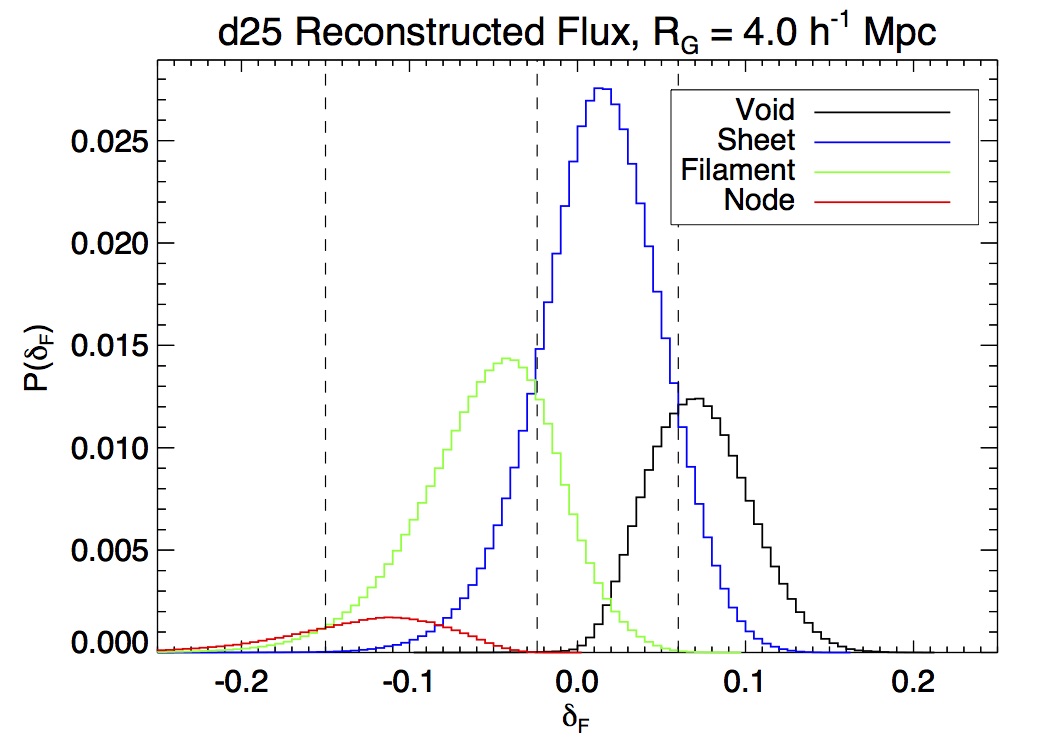} \\
\includegraphics[width=0.49\textwidth]{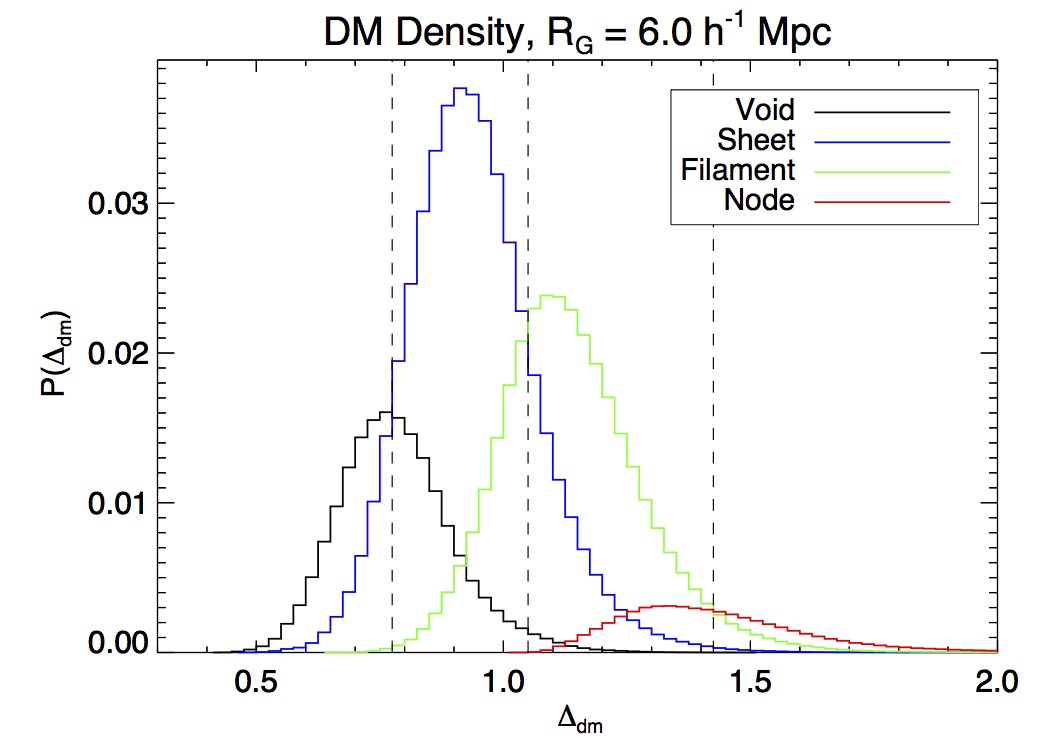}
\includegraphics[width=0.49\textwidth]{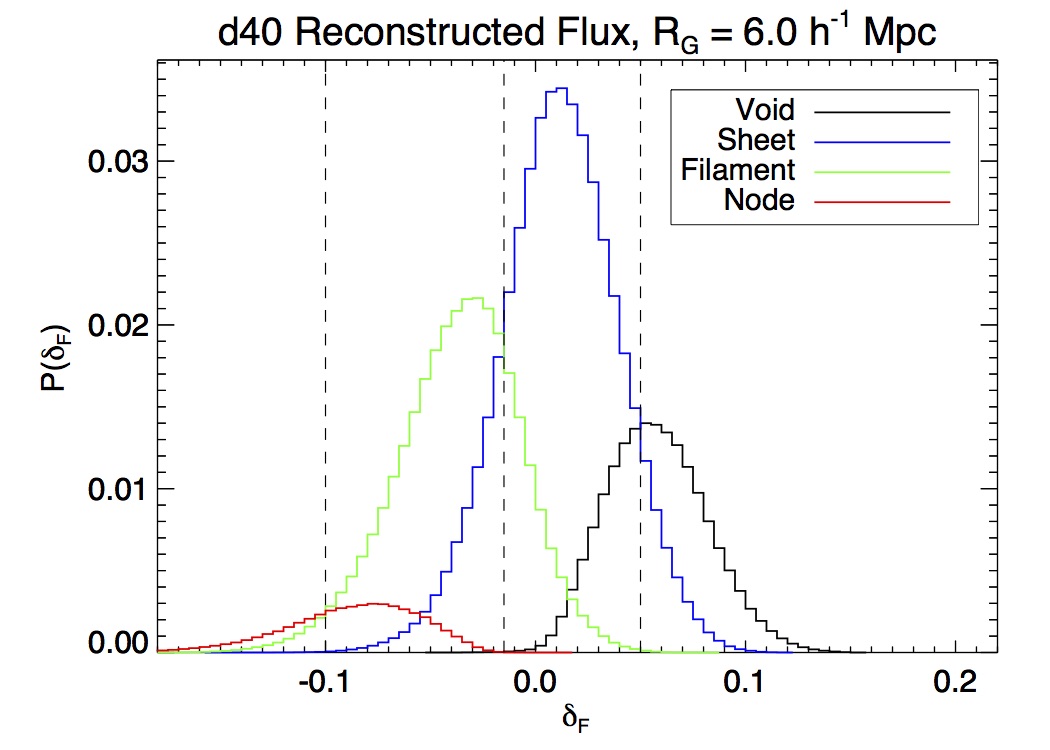}
\caption{\label{fig:pdf}
Distributions of map values corresponding to the cosmic web classifications identified through the DM overdensity (left) and reconstructed \lya\ flux maps (right), in all cases evaluated in $(1\,\hMpc)^3$ bins.
The top row show the density and \dsmall maps smoothed with a $R_G=4\,\hMpc$ Gaussian, 
while bottom row is for the density and \dmid maps smoothed with $R_G=6\,\hMpc$. The distributions are normalized such that the total area of all the distributions for a given map sum to unity. In all cases, there is significant overlap in the overdensity or flux fluctuations
assigned to the different components of the cosmic web --- the vertical-dashed lines delineate transition values
at which different components dominate. Note that negative $\delta_F$ flux values correspond to larger overdensities, and also that all the ordinate axes have different scales.}
\end{figure*}

As in other studies \citep[e.g.,][]{hahn:2007,alonso:2015,eardley:2015} we find considerable overlap in the overdensities or flux fluctuations assigned to the different cosmic web components (Fig.~\ref{fig:pdf}).
In the $R_G=4\,\hMpc$ DM map, we find that the transitional overdensity values at which voids, sheets,  filaments, and nodes successively dominate are at $\Ddm \approx [0.73, 1.1, 1.7]$. 
The corresponding boundaries for the \dsmall flux maps are $\delta_F = [0.06, -0.024, -0.15]$ 
(recall that negative $\delta_F$ corresponds to higher overdensities). 
These flux values can be converted to an equivalent DM overdensity using the  
conversion between the \dsmall flux and smoothed DM overdensity presented in \citet{lee:2016}:
\beq
\label{eq:dm_flux}
  \Ddm \approx 5.502\; \delta_F^2 - 3.681\; \delta_F + 0.947.
\eeq
When we convert the void/sheet, sheet/filament and filament/node boundary values from the
flux map with this expression, we get $\Ddm \approx [0.75, 1.04, 1.62]$.
The transition between the cosmic web element in the flux map occur at 
the nearly the same equivalent overdensities as in the DM maps.
We find a similar correspondence between the transitional values for the cosmic web in the
\dmid map and density field smoothed to $R_G=6\,\hMpc$.

Interestingly, tuning the free parameter in each map ($\lambda_{\rm th}$ for the dark matter and $\lambda_{{\rm th},F}$ for the flux) to match the void fraction in S15a produces essentially identical volume fractions of 
the other cosmic web components in both the DM and flux maps (Table~\ref{tab:basicfrac}).
In both the DM and flux maps with $R_G=4\,\hMpc$, approximately $[19\%, 49\%, 28\%, 4\%]$ of the volume are occupied by voids, sheets, filaments and nodes while the corresponding fractions in the $R_G=6\,\hMpc$ maps are $[18\%, 46\%, 31\%,  5\%]$.


By eye, there is generally a good match between the cosmic web components identified in the
tomographic flux maps \textit{vis-\`a-vis} that from the equivalently-smoothed DM map.
The voids and sheets appear to yield the best agreement, in part because they are extended structures that are more easily resolved by even relatively sparse IGM tomographic surveys.
The filaments, which occupy a smaller volume fraction, are smaller-scale structures and those
recovered from the tomography do not always have the same topology as in the underlying DM field, but nevertheless the overall match is good. The nodes occupy small isolated patches within the volume, and appear to be the worst-recovered of the cosmic web classifications.
However, in most cases the nodes in the flux maps
can be associated with DM nodes in the same vicinity, albeit with positional offsets and different sizes.

To further quantify the matching, we evaluate the volume overlap for each component recovered from the mock tomographic maps and underlying DM field.
For the voids, sheets, filaments and nodes in the \dsmall flux map, we find overlaps of approximately $[68\%, 72\%, 66\%, 52\%]$, respectively, with those from the matched DM map. 
The coarser \dmid tomographic map also provides a good, albeit slightly worse, cosmic web recovery: we find $[60\%, 67\%, 64\%, 50\%]$ overlap with voids, sheets, filaments, 
and nodes, respectively, recovered from the underlying DM.
Another method of comparison \citep{eardley:2015} is to ask what fraction of the total volume is classified to within $[-1,0,+1]$ eigenvalues?  For our \dsmall map we find these fractions to be $[15\%, 69\%,15\%]$, with only 1\% of the volume misclassified by $\pm 2$ eigenvalues.

Given the level of finite sampling and pixel noise modeled in the tomographic maps, we find this level of overlap highly encouraging.
For comparison, \citet{eardley:2015} applied a similar deformation tensor-based cosmic web classification scheme on a mock-up of the GAMA galaxy redshift survey at $0.04<z<0.26$, and found that $[10\%, 75\%, 15\%]$ of their map volume was classified to within $[-1, 0, +1]$
eigenvalues.
This is remarkable since it shows that \lya\ forest tomographic surveys using existing or near-future instrumentation should be able to classify the $z\sim 2.5$ cosmic web to an accuracy nearly as high as $z\sim 0.1$ galaxy surveys targeting sub-luminous galaxies at high number-densities. 

\subsection{Effect of Finite Survey Geometries}

So far, we have applied the deformation tensor analysis to maps covering the full $L=256\,\hMpc$ simulation volume with periodic boundary conditions. 
This is in contrast with real observational volumes, which are obviously non-periodic, and which may be smaller. 
Since the deformation tensor is a non-local function of the flux (or density) we expect
the classification to be affected by edge effects at the survey boundaries and 
to be degraded compared to this ideal case.
Moreover, since the Fourier transform and its inverse (Equation~\ref{eq:deform})
implicitly assume periodic boundary conditions, one might
also expect errors to arise from the non-periodicity of realistic survey volumes.

We investigate the impact of finite survey boundaries by analyzing sub-volumes from the \dsmall simulated map,
 at scales comparable to upcoming observations.
Since our $L=256\,\hMpc$ box size is, coincidentally, roughly equivalent to the line-of-sight distance probed by existing IGM tomography surveys\footnote{For example, the CLAMATO pilot map reported in \citet{lee:2016} spanned $260\,\hMpc$ along the line-of-sight between $2.2<z<2.5$.} we subsample only in the transverse ($xy$) dimensions to simulate different survey areas covered by \lya\ forest tomography;
all the sub-volumes will still span the full $256\,\hMpc$ along the line-of-sight ($z$-dimension).

\bfig
\includegraphics[width=0.51\textwidth,clip=true,trim=20 0 0 0]{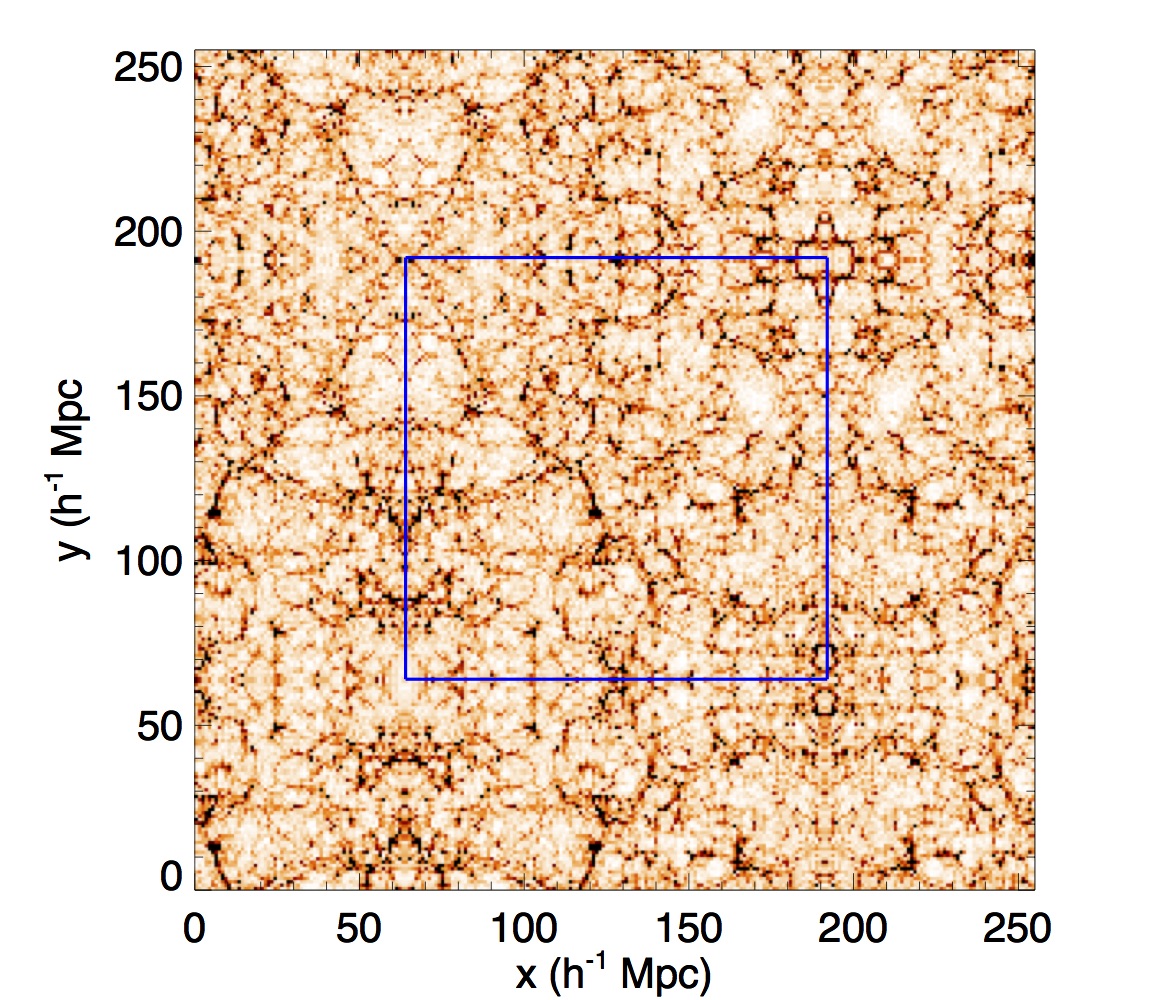}
\caption{\label{fig:padrefl} Illustration of our reflective padding procedure:
a $(128\,\hMpc)^2$ density sub-volume from the simulation (enclosed in blue box) 
is embedded in a larger super-grid, and padding outside of the actual data is
introduced by reflecting the data along the $x$ and $y$ boundaries. Note that this is just an 
illustrative example: for our actual tests we embed the test sub-volumes in a $512^3$ super-grid.
}
\efig

We will test the cosmic web recovery on 3 different finite survey areas: $(16\,\hMpc)^2$, $(64\,\hMpc)^2$, and $(128\,\hMpc)^2$. The smallest survey area, $(16\,\hMpc)^2$, is slightly larger than, but 
roughly equivalent, to the
$14\,\hMpc \times16\,\hMpc$ map area covered by the recent pilot observations of \citet{lee:2016}.
The $(64\,\hMpc)^2$ area corresponds to roughly one square degree on the sky at these redshifts, which the full
CLAMATO survey aims to cover over the next several years. 
Further in the future, massively-multiplexed spectroscopic surveys on $8-10\,$m class telescopes will
easily cover areas of several square degrees, e.g.\ the Galaxy Evolution component of 
the Subaru Strategic Program for Subaru-PFS \citep{takada:2014}
will observe multiple regions across the sky, 
each spanning $4-5$ square degrees --- the $(128\,\hMpc)^2$ map will approximate
such a wider survey geometry for each of these chunks. The size of these transverse areas, relative to the full simulation box, are illustrated in the lower-right panels of Fig.~\ref{fig:cosmicweb1} and Fig.~\ref{fig:cosmicweb2}.
Purely by eye, one might expect an aliasing effect from
the survey geometry of the smallest $(16\,\hMpc)^2$ area,
which is not much larger than the characteristic scales spanned by the 
cosmic web structures at the scales probed by IGM tomography.

\bfig
\includegraphics[width=0.49\textwidth,trim=10 0 0 0, clip=true]{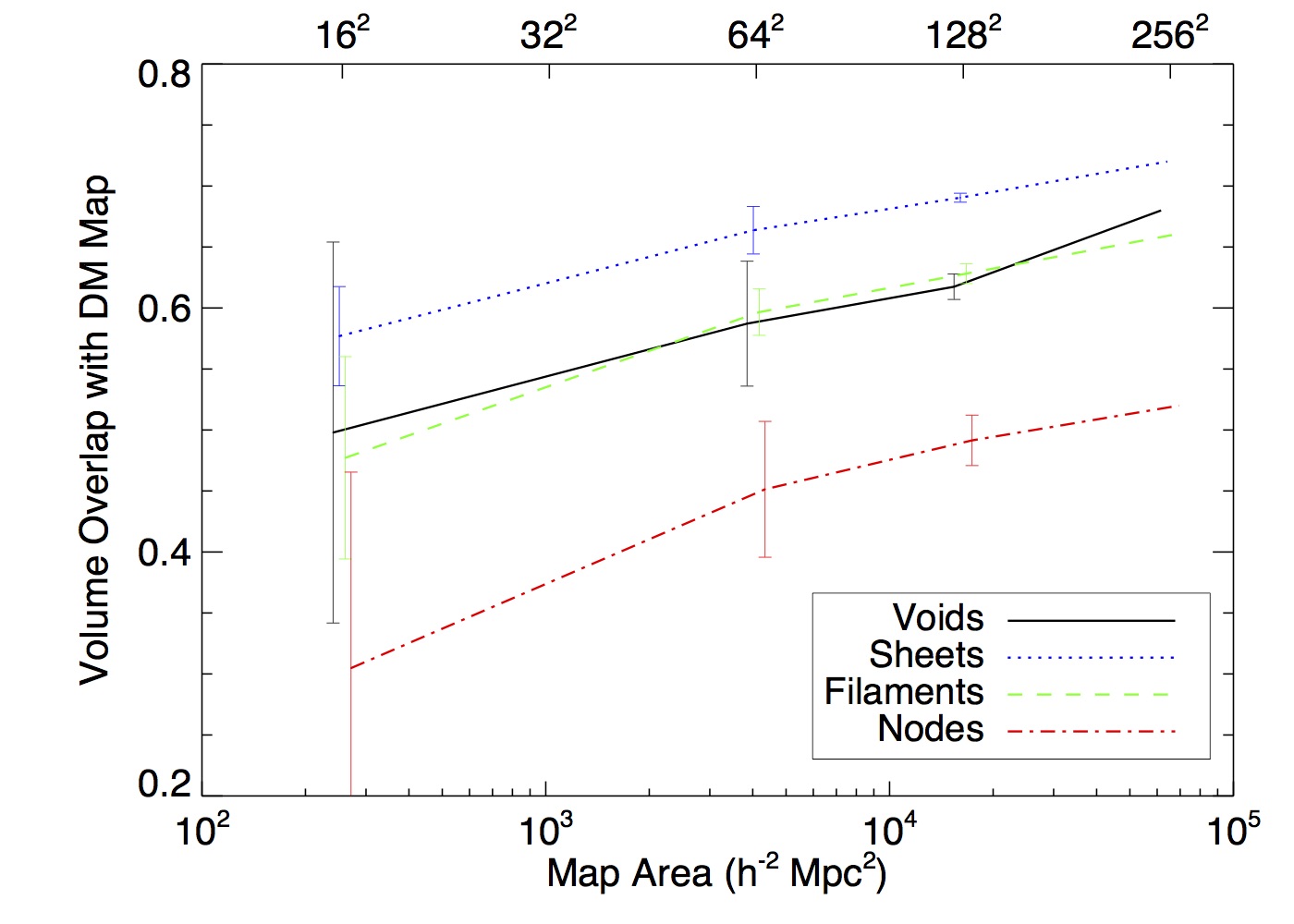}
\caption{\label{fig:subvol_overlap}
Volume overlap fractions of voids, sheets, filaments and nodes in the $z=2.5$ cosmic web recovered
from subvolumes of the \dsmall tomographic flux map, compared with those
in the underlying full DM map with $R_G = 4\,\hMpc$. In each subvolume, the subsampling
is carried only in the transverse $xy$ plane, while the $z$-dimension spans
the full $256\,\hMpc$ of the simulation box. This is carried out over [36,16,4]
realizations of sub-volumes with areas of [$(16\,\hMpc)^2,\, (64\,\hMpc)^2,\, (128\,\hMpc)^2$] --- the error bars denote the standard deviation of the 
realizations.
The recovery of the cosmic web is degraded
in map areas significantly smaller than $\sim (60\,\hMpc)^2$, equivalent to approximately one square degree
at these redshifts. 
}
\efig

To mitigate the edge effects, some form of padding outside the
actual data volume is desirable. Several methods exist to carry out this 
padding in the context of Fourier analyses of large-scale structure
, see e.g., \citet{press:1992,blake:2010,chiang:2013b}. 
We choose follow the method described in \citet{eardley:2015}, 
which involves padding the region outside of the map by reflecting half of the data along each edge.
To allow this, we define a $512^3$ super-grid with $1\,\hMpc$ cell size, and embed each of the test
sub-volumes in the middle of the super-grid. 
We then perform the reflective padding across boundaries in all three dimensions 
(illustrated in Figure~\ref{fig:padrefl}), 
since the $512^3$ super-grid was
chosen to be large enough to allow the reflective padding even along the long $z$-dimension. 
Beyond the
reflection regions up to $L/2$ beyond the boundaries, we pad the super-grid with zeros. 
The deformation tensor evaluation and eigenvalue
analysis is carried out as described previously, 
and using the same eigenvalue thresholds as before.

After classification, the input sub-volume is then extracted from the padded super-grid, 
and the cosmic web elements therein compared with those
in the corresponding regions of the full periodic DM map. 
As before, we simply compute the volume overlap fractions between each cosmic web
component derived from the mock survey sub-volumes, and the `true' cosmic
web in the equivalent DM volume at the same smoothing scale. One could quantify this
using more sophisticated metrics, but to ease comparison with \citet{eardley:2015} we
persist with the volume overlap. In any case, the mock surveys presented here are
merely rough strawmen at what could be achievable in the near future, and a more 
detailed analysis should be carried out with more realistic mock surveys once the
actual survey parameters are better defined.

We average the analysis over a number of independent sub-volume realizations corresponding to 
each survey size that we test:
we analyzed [36, 16, 4] realizations with areas of [$(16\,\hMpc)^2,\, (64\,\hMpc)^2,\, (128\,\hMpc)^2$], 
respectively.
The results are summarized in Fig.~\ref{fig:subvol_overlap}. As expected, smaller maps lead to a degraded recovery of the cosmic web: within the
$(16\,\hMpc)^2$ sub-volume, the filaments and nodes are recovered with only $\approx 45\%$ and $\approx 26\%$
volume overlap with the true DM map, while the voids and sheets are recovered with
 $< 60\%$ volume overlap. There is also considerable scatter of
 up to $\sim 15\%$ rms between realizations, particularly in the recovery
 of the voids and nodes. This is a clear degradation compared to the recovery fraction of
 $[72\%, 66\%, 52\%]$ for the sheets, filaments and nodes in the full periodic
 $L=256\,\hMpc$ flux map.
By going to observed areas of order a square degree ($(64\,\hMpc)^2$ area), 
the volume overlap with the DM is improved
significantly to $\approx [55\%, 66\%, 59\%, 49\%]$ for the voids, sheets, filaments, and nodes, respectively,
in the $(64\,\hMpc)^2$ map. The scatter between sub-volumes
is also significantly reduced to no more than $\sim 5\%$ rms. 
Finally, the $(128\,\hMpc)^2$ sub-volumes achieve nearly the nearly the same
level of cosmic web recovery as the full volume with period boundary conditions. This indicates that future surveys using ultra wide-field spectrographs, such as Subaru-PFS, should target contiguous fields of several square degrees in order to minimize effects from the finite boundary conditions.

\section{Halo Abundances in the Cosmic Web}

There is some evidence to suggest that the manner in which halos form, and some of their properties, depend upon their location within the cosmic web \citep[e.g.][]{hahn:2007,libeskind:2012,forero-romero:2014}.
In as much as galaxy properties are determined by the formation history and properties of the dark matter halos in which they reside, and their nearest neighbors \citep{porter:2008}, we may expect to see a dependence of galaxy properties on cosmic web location as well.  It is plausible, though by no means demonstrated, that galaxy properties may depend on the larger scale (gaseous) environment in which the galaxy forms.

At lower redshifts, a major determinant of galaxy properties appears to be halo mass (see, e.g., \citealt{abbas:2006, skibba:2006, tinker:2008}; although some properties depend upon formation history, e.g.~\citealt{reed:2007, yan:2013, hearin:2015}), and it has been shown in simulations that the major determinant of the halo mass function appears to be local density (or the trace of our Hessian matrix; \citealt{forero-romero:2014,alonso:2015}).  This is expected in a theory based on Gaussian statistics within the linear regime, where the halo mass function varies solely due to the underlying density field and there is no coupling to tidal forces \citep{alonso:2015}.  

Here we make a preliminary investigation of
 the relationship between the geometric cosmic web classification and local density using the tomographic flux maps, and the manner in which the halo mass function\footnote{We defer a detailed examination of other halo properties to a future publication, where we would also like to investigate gas flows and halo fueling using hydrodynamical, rather than dark matter-only, simulations.} depends on cosmic web classification, controlling for local density, at $z\sim 2.5$.  We find that environment and local density are correlated, and at fixed density there is little dependence of halo multiplicity on environment even at these early times.

We use a friends-of-friends halo catalog at $z\simeq 2.5$ for our investigations.
First, we define the regions in the simulation volume corresponding to either 
the overdensity range or cosmic classification, and then compute the multiplicity function of the encompassed halos:
\beq
f(\ln M)\, \mathrm{d}\!\ln M = \frac{M}{\bar{\rho}} \frac{\mathrm{d}n}{\mathrm{d}\!\ln M} \mathrm{d}\!\ln M,
\eeq
where $M$ is the (FoF) halo mass, $n$ is the number density, and $\bar{\rho}$ is the mean mass density defined within the volume of the overdensity or cosmic web cut.
 
\begin{figure*}
\includegraphics[width=0.49\textwidth]{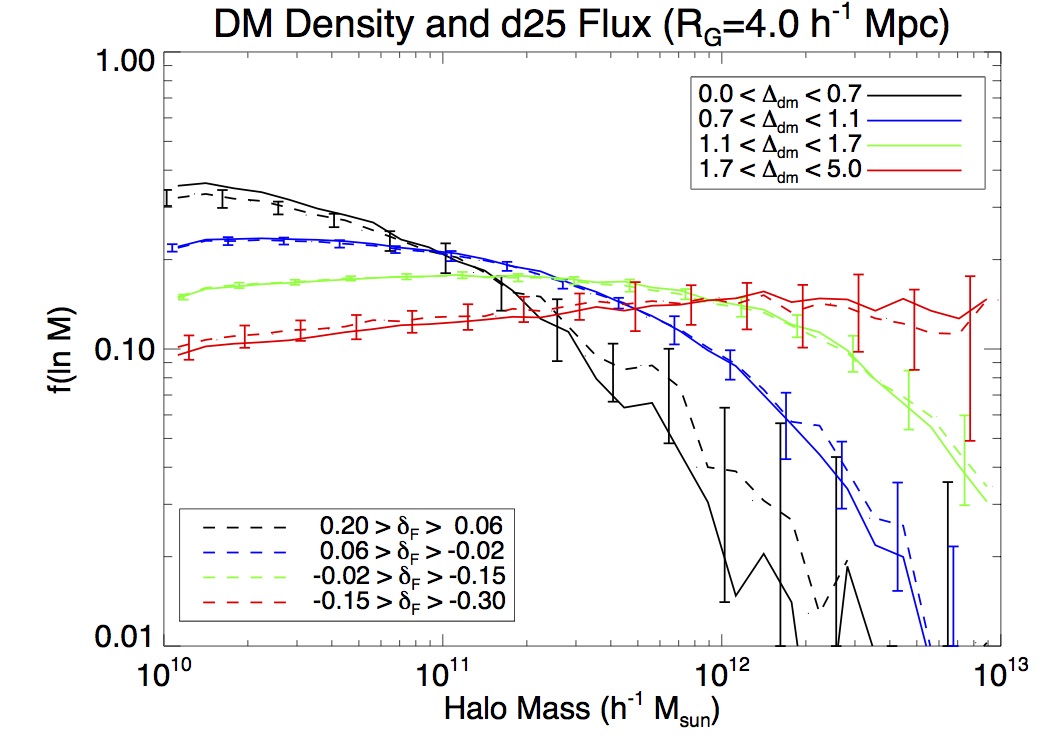} 
\includegraphics[width=0.49\textwidth]{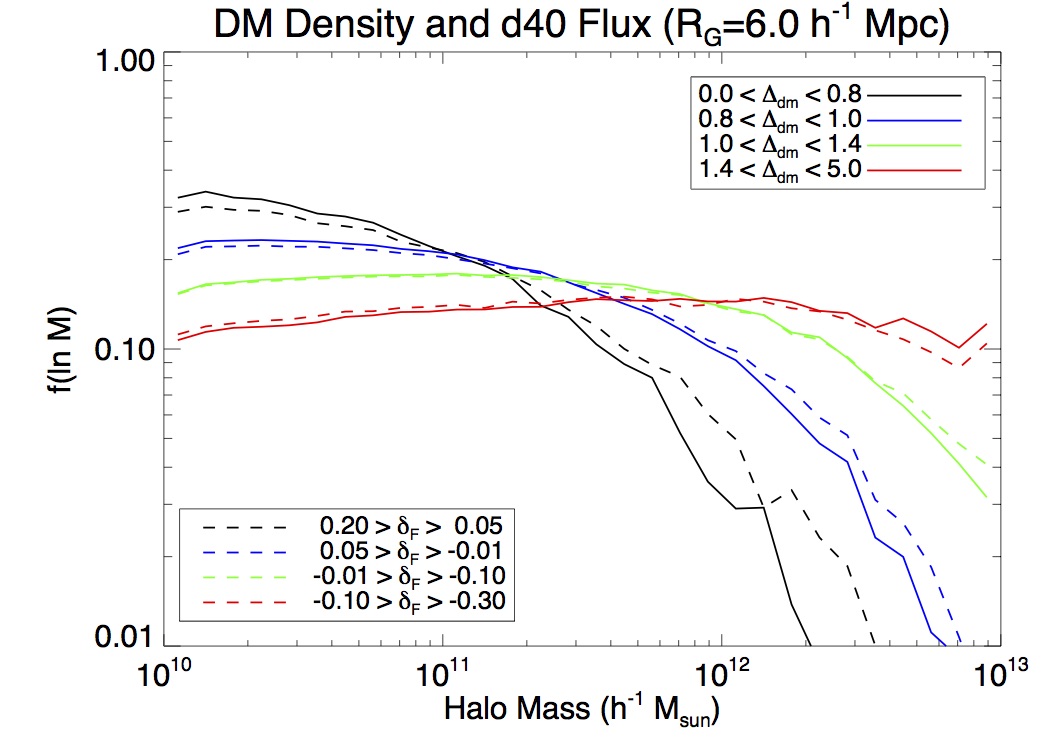}
\caption{\label{fig:massfn_den}
Halo mass multiplicity functions evaluated as a function of density environment in maps 
smoothed by $R_G=4\,\hMpc$ (left) and $R_G=6\,\hMpc$ (right). In each case, the solid lines
show the multiplicity functions measured from halos residing in different DM density ranges, while dashed lines show
those for several ranges in reconstructed \lya\ flux, $\delta_F$. The DM density and flux ranges were chosen
to be those at which different cosmic web classifications dominate (Fig.~\ref{fig:pdf}). The error bars on the left
panel show the standard deviation from evaluating CLAMATO-like ($(64\,\hMpc)^2$) 
flux subvolumes.
} 
\end{figure*}

\begin{figure*}
\includegraphics[width=0.49\textwidth]{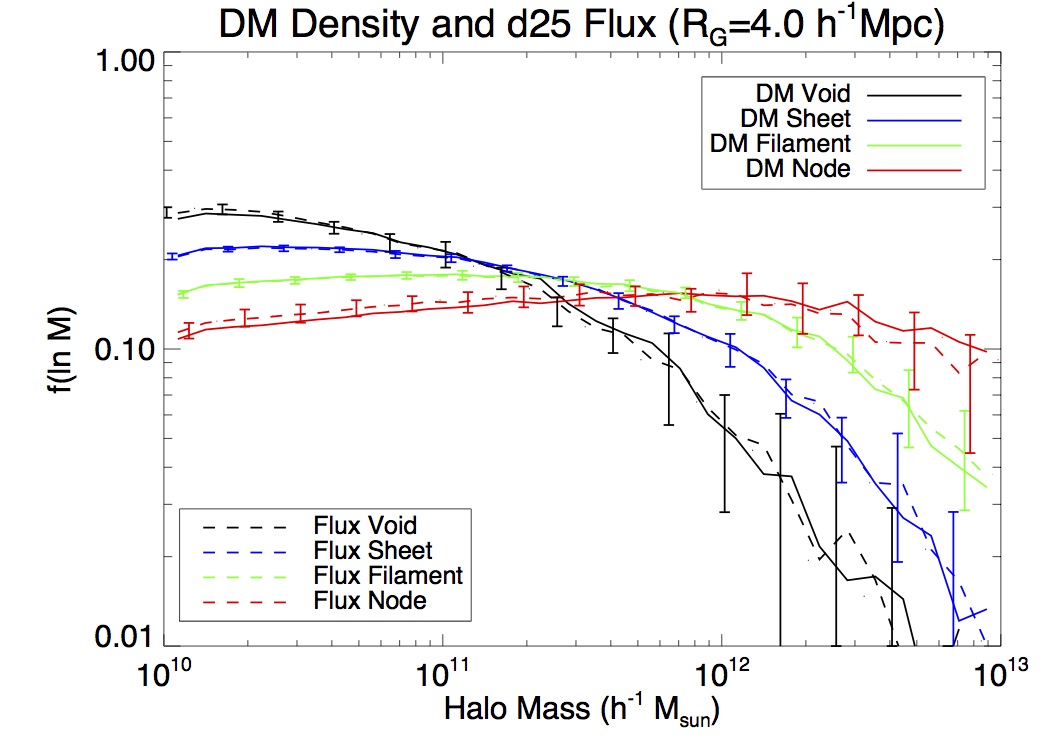} 
\includegraphics[width=0.49\textwidth]{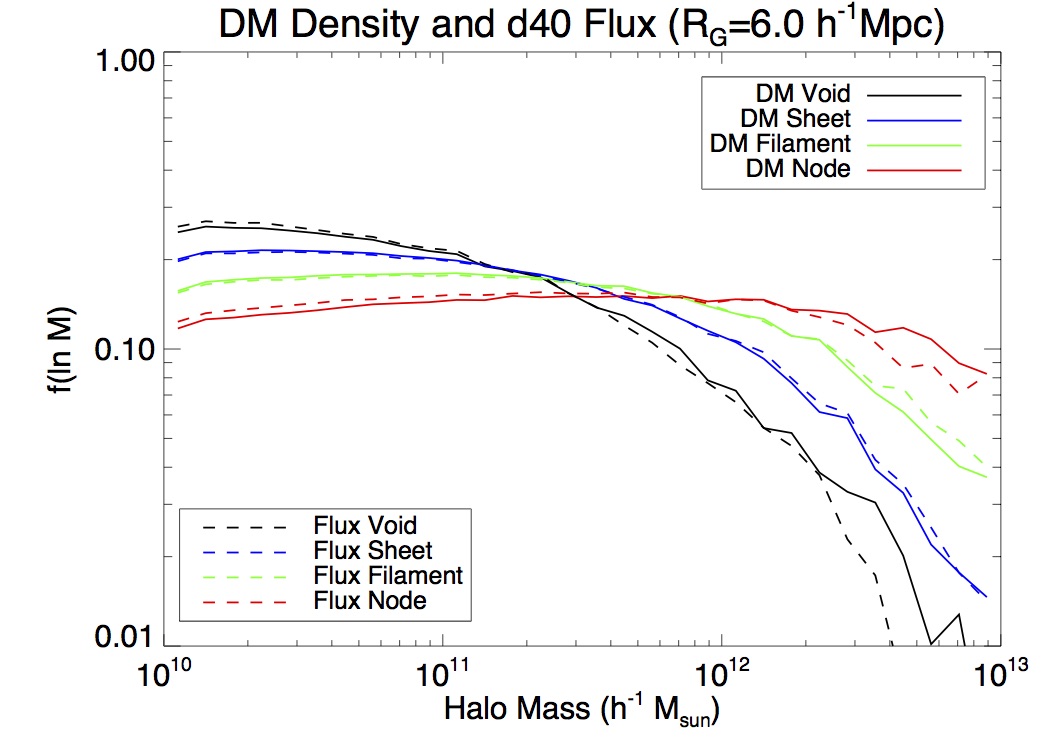}
\caption{\label{fig:massfn_cw}
Same as Fig.~\ref{fig:massfn_den}, but evaluated as a function of geometric cosmic web environment recovered from each map.
} 
\end{figure*}

In Figure~\ref{fig:massfn_den}, we show the multiplicity functions split by both DM overdensity and 
reconstructed \lya\ flux in the \dsmall map, in which the ranges were chosen to be between the transitional values 
between different cosmic web components (vertical-dashed lines in Fig.~\ref{fig:pdf}). 
As expected, we see large differences in the halo multiplicity: halos residing in low-density environments
tend to have lower halo masses, while high-density regions harbor relatively massive halos.
In other words, the local density (``environment'') can act as a proxy for halo mass, as has been studied using lower-$z$ galaxy environment estimators \citep[e.g.,][]{haas:2012}.
Interestingly, the multiplicity functions measured in the DM density maps smoothed by the two different smoothing scales ($R_G=4\,\hMpc$ and $R_G=6\,\hMpc$) differ by only $\sim 10\%$.
This suggests that a cosmic web classification would be useful to define consistent overdensity thresholds 
across different surveys sampling various smoothing scales, even if (as we shall see) there is 
little explicit dependence of the halo abundance on the geometric cosmic web.

It is also clear from Fig.~\ref{fig:massfn_den} that in terms of differentiating environments
with different halo abundances, the reconstructed \lya\ flux is a good proxy for the DM overdensity at equivalent smoothing scales. This is
even more impressive since we did not select the $\delta_F$ ranges arbitrarily: they were chosen from the cosmic web
transitions defined self-consistently from the flux maps 
(right panels in Fig.~\ref{fig:pdf}) independently
of any information about the DM multiplicity functions
--- which were themselves defined via as the DM cosmic web
transitional values.

We find a slight bias in the multiplicity functions
measured in the lowest-flux volumes relative to the lowest overdensity bin in the DM.
To quantify this further in terms of upcoming surveys, for the \dsmall flux map we evaluate the multiplicity functions within a
CLAMATO-like survey volume:
we make $(64\,\hMpc)^2$ cuts in the transverse plane and plot the standard 
deviation of the resulting halo abundances as the error bars in 
Fig.~\ref{fig:massfn_den}. We did not estimate errors for the \dmid box since the total survey volume for 
a $\sim 20\,\sqdeg$ Subaru-PFS survey would be similar to our entire simulation volume.

We find that within the observational uncertainties expected from the $\sim 1\,\sqdeg$ CLAMATO survey (error bars in left panels of Figs~\ref{fig:massfn_den} and \ref{fig:massfn_cw}), 
 the density-dependent halo multiplicity functions defined on the
flux maps are consistent with those from the DM density.
This indicates that IGM tomography surveys will be 
enable unique insights into galaxy evolution at $z\simeq 2.5$, 
allowing the \lyaf\ flux to act as a proxy for environmental density \citep[and hence halo mass,][]{haas:2012}. 
While it would be extremely challenging to measure enough
spectroscopic redshifts of coeval galaxies to directly compute halo abundances, 
IGM tomographic maps will allow us to identify regions with different underlying halo abundances.
This information can then be related to observed galaxy properties from even limited samples of coeval galaxies.

We also evaluate the halo multiplicity as a function of the geometric cosmic web classification, 
the results of which are shown in Fig.~\ref{fig:massfn_cw} for both the DM and flux maps at 
$R_G=4\,\hMpc$ and $R_G=6\,\hMpc$. There are also clear differences in the halo multiplicities
seen in the different cosmic web components within a given map, but at a somewhat lower level
than those seen in the overdensity cuts (Fig.~\ref{fig:massfn_den}). 
This reduced contrast is primarily because the halo abundance depends primarily
on the overdensity environment rather than explicitly on the cosmic web classification,
therefore evaluating $f(\ln M)$ within each cosmic web component actually samples a relatively broad range of overdensities than when sampling explicitly by overdensity (see Fig.~\ref{fig:pdf}).

By comparing halos that reside in the same overdensity environment but different cosmic 
web classifications, we can test for any explicit dependence on the geometric cosmic
web environment. In the DM map with $R_G=4\,\hMpc$ smoothing, 
we select halos that reside in regions with $0.82 < \Ddm < 0.92$. This is a range
that is dominated by sheets, but simultaneously includes the low- and high-mass tails 
of filaments and
voids, respectively (c.f.~top-left panel of Fig.~\ref{fig:pdf}). To remove any relative bias in the
overdensity distribution within this range, 
we match the halo counts to
within bins of $\Ddm = 0.005$ for all the cosmic web components. Due to this requirement, 
we are limited to evaluating only a total of 10,000 halos for each component, resulting
in somewhat noisy multiplicity functions.
We also carry out the same exercise for the \dsmall flux maps by selecting halos found in regions with
$0.012 < \delta_F < 0.036$, which is approximately equivalent
to the aforementioned $0.82 < \Ddm < 0.92$ range according to Eq.~\ref{eq:dm_flux}.

\bfig
\includegraphics[width=0.49\textwidth]{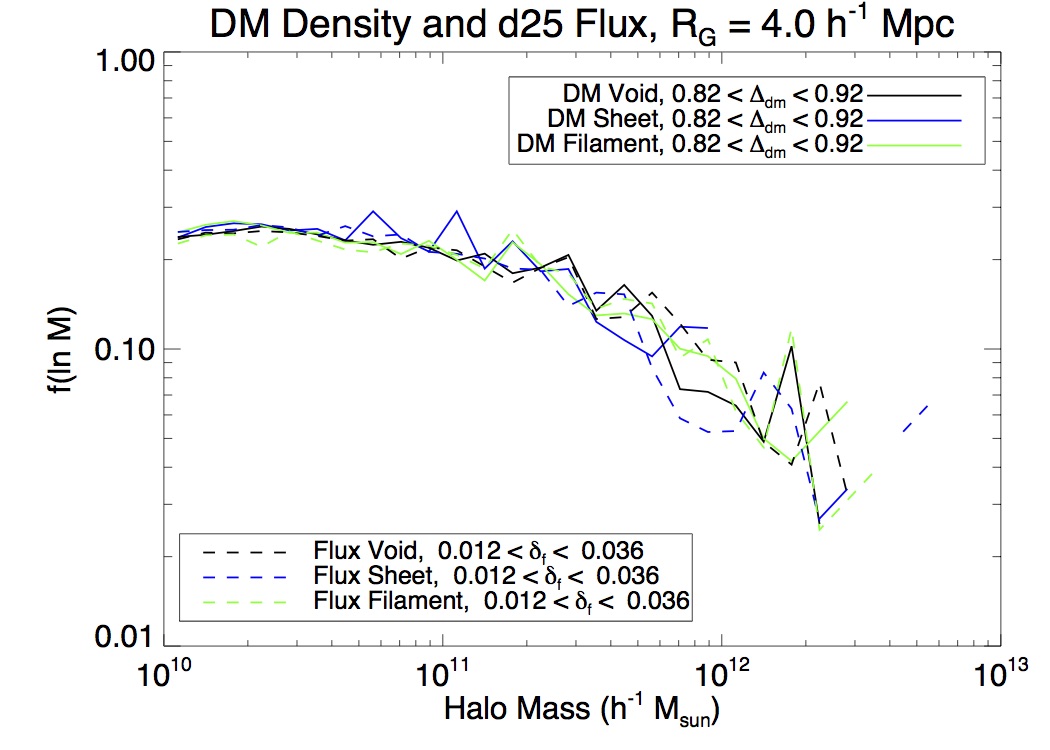}
\caption{\label{fig:massfn_fixden}
Halo multiplicity functions evaluated in different geometric cosmic web environments, but at
fixed DM density (solid curves) or fixed \lya\ flux in the \dsmall map(dashed lines). 
In both cases, the maps have been smoothed with a $R_G = 4\,\hMpc$ Gaussian.
There is no significant difference in the halo multiplicity with geometric cosmic web
environment when the density is held fixed.}
\efig

The resulting halo multiplicity functions are shown in Fig.~\ref{fig:massfn_fixden}.
While the curves are somewhat noisy, there is no significant difference in halo abundance as a function of cosmic web classification when the density is held fixed, even when contrasting voids with filaments. This is consistent with the low-redshift behavior found by e.g.~\citet{alonso:2015}, who also found no explicit cosmic web dependence on scales of several Mpc.

However, when considering halos within the cosmic web defined on the reconstructed
\lya\ flux, we find a slight (several percent) tilt in the relative
halo multiplicity when going from voids to filaments. This is caused by a slight bias
in the flux filament halos toward higher \Ddm\ than would be expected from the mean
$\Ddm$-$\delta_f$ relationship (Eq.~\ref{eq:dm_flux}).
This bias is possibly due to the different redshift-space distortions in the flux and DM fields and could potentially be used to improve the recovery of the underlying DM fields from IGM flux maps, but we defer this to subsequent papers.

\section{summary and conclusion}

In our current cosmological model, all objects form and evolve as part of the cosmic web of large-scale structure.  When smoothed on Mpc scales, the cosmic web may be classified into voids, sheets, filaments and nodes and as the web evolves material flows out of voids, through sheets and along filaments into nodes.
It remains an area of active investigation whether, and in what manner, the location of cosmological objects within the web affects their evolution and properties.

The cosmic web has been studied at low redshift using spectroscopic samples of galaxies 
with large number densities, but maintaining Mpc resolution over cosmological volumes with such surveys becomes progressively more difficult at higher redshift.
Beyond $z\simeq 2$ a new technique can be applied for mapping large-scale structure 
and the cosmic web: Ly$\alpha$ forest tomography.
By studying Ly$\alpha$ absorption in the spectra of distant QSOs and galaxies 
with small transverse separations, it is 
possible to map cosmologically representative volumes with Mpc-scale fidelity and 
thus classify regions within the cosmic web.

We have found that in our simulations the cosmic web classification in \lyaf\ flux maps --- reconstructed
from realistic mock data --- is in excellent agreement with that determined directly from the underlying dark matter for volumes and resolutions comparable to the ongoing CLAMATO survey.
This offers a promising route, with current facilities, to studying galaxies and AGN within the context of the cosmic web during ``cosmic noon'', when we expect vigorous star-formation activity, AGN activity and gas inflows (possibly along filaments).

The primary observational limitation to classifying structures within the cosmic web comes directly from the non-local nature of its definition.  For a survey which is too small, boundary effects dramatically affect the accuracy of the classification.  Since structures in the cosmic web span tens of Mpc, the survey must cover $1\,{\rm deg}^2$ in order to return reliable classifications. Once a survey reaches this size, however, the classifications are
nearly as reliable as those obtained from deep spectroscopic surveys of the local Universe.

This paper establishes that a reliable cosmic web classification is possible with Ly$\alpha$ 
forest tomography on existing facilities with relatively modest investments of time. 
We also made a preliminary investigation
into halo properties as a function of the cosmic web, 
which showed that the high-$z$ multiplicity function of halos was largely insensitive to where in the cosmic web a halo lay, depending primarily on local density.  This suggests that only galaxy properties which depend on properties other than halo mass (e.g.~formation history, spin, etc.) are likely to correlate with geometric environment even at high $z$.  In future work we will investigate the predictions of models of galaxy and AGN formation, and gas flows in hydrodynamic simulations, for more detailed dependence on geometric environment during this active era in galaxy evolution. 
Nevertheless, we have shown that the IGM tomographic flux is an excellent proxy for the scalar
density environment at $z\sim 2.5$, therefore the combination of upcoming 
maps from the CLAMATO survey with coeval galaxy samples will likely yield exciting new results.

\bibliographystyle{apj}

\acknowledgements{We thank Joanne Cohn for stimulating and helpful discussions, and Casey Stark for making available his simulation products.
KGL acknowledges support for this work by NASA through Hubble Fellowship grant HF2-51361 awarded by the Space Telescope Science Institute, which is operated by the Association of Universities for Research in Astronomy, Inc., for NASA, under contract NAS5-26555.

\bibliography{lyaf_kg,apj-jour,lss_galaxies,my_papers}

\end{document}